\documentclass[11pt]{article}

\begin{document}

\pagestyle{empty}

\hyphenation{brems-strahlung had-rons im-p-or-t-ant
             mis-iden-ti-fy mis-iden-ti-fied mis-iden-ti-fi-ca-tion}

\def\={\hbox{$\;$}}
\def\undertext#1{$\underline{\hbox{#1}}$}
\def\olddfrac#1#2{\frac{\displaystyle#1}{\displaystyle#2}}
\def\dfrac#1#2{{\displaystyle\frac{#1}{#2}}}
\def\sfrac#1#2{{\displaystyle{#1}/{#2}}}
\def\al{\alpha}
\def\bal{\mbox{\boldmath$\alpha$\unboldmath}}
\def\bold#1{\mbox{\boldmath$#1$\unboldmath}}
\def\bV#1{\mbox{$\bold V\mkern-5mu_{#1}$}}
\def\BV#1{\mbox{$\bold V^{-1}_{\mkern-8mu#1}$}}
\def\p{\phantom{-}}
\def\e{$\equiv$}
\def\h{\hskip-3mm}
\def\BaBar{{\sc BaBar}}
\renewcommand{\thefootnote}{\fnsymbol{footnote}}


\begin{flushright}
{\small
SLAC--PUB--8650\\
UMS/HEP/00/09\\
October 2000\\}
\end{flushright}

\vspace{.8cm}

\begin{center}
\baselineskip6mm{\Large\bf
  Kinematic Fit for the Radiative Bhabha Calibration
of \BaBar's Electromagnetic Calorimeter\footnote{Work supported 
by Department of Energy contract DE--AC03--76SF00515 and 
Department of Energy grant DE-FG02-91ER40622.}}
\end{center}

\vspace{1cm}

\begin{center}
Johannes M. Bauer\\ \vspace{2mm}
Department of Physics and Astronomy, 
University of Mississippi, 
University, MS 38677, U.S.A.\\
Stanford Linear Accelerator Center, 
Stanford University,
Stanford, CA 94309, U.S.A.
\end{center}

\vfill

\begin{center}
{\large\bf
Abstract }
\end{center}

\begin{quote}
For the radiative Bhabha calibration of \BaBar's electromagnetic
calorimeter, the measured energy of a photon cluster is being compared
with the energy obtained via a kinematic fit involving other
quantities from that event.  The details of the fitting algorithm are
described in this note, together with its derivation and checks that
ensure that the fitting routine is working properly.
\end{quote}

\vfill


\newpage
\pagestyle{plain}

\section{Introduction}

Radiative Bhabhas can be used as one of the calibrations of the \BaBar\ 
electromagnetic calori\-meter~(EMC)\null.  Radiative Bhabha events
($e^-e^+\rightarrow e^-e^+\gamma$) deposit photons over a large energy
range everywhere in the calorimeter.  If the momenta of the incoming
and outgoing electrons and positrons, as well as the photon's angular
position are known, the photon energy can be obtained via a kinematic
fit.  This fit results in an absolute measurement of the photon energy
which then can be compared to the measured photon energy to obtain
calibration constants.

The radiative Bhabha module is part of \BaBar's Online Prompt
Reconstruction~(OPR) executable.  Initial cuts select good electrons,
positrons, and photons.  Then all possible combinations of triplets
(one electron, one positron, one photon) are formed.  Each triplet is
sent to the fitting routine to calculate its $\chi^2_{\rm est}$, the
``estimated $\chi^2$''.  The triplet with the lowest~$\chi^2_{\rm
est}$ is then submitted to the full kinematic fit which returns, among
other quantities, the fitted photon energy $E_{f\gamma}$ and the error
matrix of the fitted quantities.  The ratio $E_{\rm meas}/E_{f\gamma}$
is later used to calibrate the calorimeter.  Note that no information
on the measured photon energy $E_{\rm meas}$ goes into the kinematic
fit or~$\chi^2_{\rm est}$.

This note is the complete documentation on the algorithm for fitting
the radiative Bhabha events for the purpose of calibrating the
calorimeter.  It~describes the whole fitting procedure: the quantities
for the kinematic fit and $\chi^2_{\rm est}$; the derivation and
formulas for $\chi^2_{\rm est}$; the derivation and algorithm for the
kinematic fit; tests to check the quality of the kinematic fit.  The
note details all formulas which go into the computer program so that
the program can be checked directly against this document.  The
derivations contain more details than needed to understand the
concept, but the details help to derive, check and recheck all
necessary formulas.  Actual results of the fitting procedure using
real data are not included in this note to keep it a pure code
documentation.

\section{Defining the quantities and constraints} 

\subsection{Measured quantities}

From the experiment come the following measurements, which shall form
the 14-dimensional vector~$\bold y$:

\medskip 
\begin{tabular}[]{lcllcllcll}
 $P_{ix-}$          &\h\e&\h $y_1$    & $P_{iy-}$        &\h\e&\h $y_2$    & $P_{iz-}$ &\h\e&\h$y_3$    &msrd momentum in $x$, $y$, and $z$ of incoming $e^-$ \\
 $P_{ix+}$          &\h\e&\h $y_4$    & $P_{iy+}$        &\h\e&\h $y_5$    & $P_{iz+}$ &\h\e&\h$y_6$    &msrd momentum in $x$, $y$, and $z$ of incoming $e^+$ \\
 $P_{ox-}$          &\h\e&\h $y_7$    & $P_{oy-}$        &\h\e&\h $y_8$    & $P_{oz-}$ &\h\e&\h$y_9$    &msrd momentum in $x$, $y$, and $z$ of outgoing $e^-$ \\
 $P_{ox+}$          &\h\e&\h $y_{10}$ & $P_{oy+}$        &\h\e&\h $y_{11}$ & $P_{oz+}$ &\h\e&\h$y_{12}$ &msrd momentum in $x$, $y$, and $z$ of outgoing $e^+$ \\
 $\theta_{o\gamma}$ &\h\e&\h $y_{13}$ & $\phi_{o\gamma}$ &\h\e&\h $y_{14}$ &           &                &                    &measured $\theta$ and $\phi$ of the photon \\
\end{tabular}
\medskip

The momenta of the incoming electron and positron and their errors are
changing run-by-run.  The errors of the incoming leptons are given as
covariance matrices:
\[
   \begin{array}{rclrcl}
      \bV {i-} &=& \left( 
            \begin{array}{ccc}
               V_{ixx-} & V_{ixy-} & V_{ixz-} \\
               V_{ixy-} & V_{iyy-} & V_{iyz-} \\
               V_{ixz-} & V_{iyz-} & V_{izz-} 
            \end{array}
                             \right) & \quad \bV {i+} &=& \left(
            \begin{array}{ccc}
               V_{ixx+} & V_{ixy+} & V_{ixz+} \\
               V_{ixy+} & V_{iyy+} & V_{iyz+} \\
               V_{ixz+} & V_{iyz+} & V_{izz+} 
            \end{array}
                                                              \right) 
   \end{array}
\]
The errors on $\bold P_{i-}$ and $\bold P_{i+}$ are assumed to be
independent.

The errors of $\bold P_{o-} = (P_{ox-},P_{oy-},P_{oz-})$ and $\bold
P_{o+} = (P_{ox+},P_{oy+},P_{oz+})$ are also assumed to be independent
from each other.  They are given in two $3\times3$ error matrices:
\[
   \begin{array}{rclrcl}
      \bV {o-} &=& \left( 
            \begin{array}{ccc}
               V_{oxx-} & V_{oxy-} & V_{oxz-} \\
               V_{oxy-} & V_{oyy-} & V_{oyz-} \\
               V_{oxz-} & V_{oyz-} & V_{ozz-} 
            \end{array}
                             \right) \quad&\quad \bV {o+} &=& \left(
            \begin{array}{ccc}
               V_{oxx+} & V_{oxy+} & V_{oxz+} \\
               V_{oxy+} & V_{oyy+} & V_{oyz+} \\
               V_{oxz+} & V_{oyz+} & V_{ozz+} 
            \end{array}
                                                              \right) 
   \end{array}
\]
The errors on $\theta_{o\gamma}$ and $\phi_{o\gamma}$ appear in the
current analysis without $\theta$-$\phi$-correlations since they were
found to be negligibly small, but we still use this $2\times2$ sub-set
of the larger $4\times4$ error matrix of the EmcCluster:
\[
   \begin{array}{rcl}
      \bV {o\gamma} &=& \left(
            \begin{array}{cc}
               V_{o\theta\theta\gamma} & V_{o\theta\phi\gamma} \\
               V_{o\theta\phi\gamma}   & V_{o\phi\phi\gamma}   
            \end{array}
       \right) 
   \end{array}
\]

All the errors can be combined in one $14\times14$ error matrix $\bV{\rm all}$.  Its
format is like this:
\[
   \bV{\rm all} = \left(
     \begin{array}{ccccc}
                \hskip-1mm\bV{i-}\hskip-1mm      &0&0&0&0 \\
       0&       \hskip-1mm\bV{i+}\hskip-1mm      &0&0&0  \\
       0&0&     \hskip-1mm\bV{o-}\hskip-1mm      &0&0   \\
       0&0&0&   \hskip-1mm\bV{o+}\hskip-1mm      &0    \\
       0&0&0&0& \hskip-1mm\bV{o\gamma}\hskip-1mm      
     \end{array}
   \right)
  =
  \left(
    \begin{array}{cccccccccccccc}
\times&\times&\times&      &      &      &      &      &      &      &      &      &      &       \\
\times&\times&\times&      &      &      &      &      &      &      &      &      &      &       \\
\times&\times&\times&      &      &      &      &      &      &      &      &      &      &       \\
      &      &      &\times&\times&\times&      &      &      &      &      &      &      &       \\
      &      &      &\times&\times&\times&      &      &      &      &      &      &      &       \\
      &      &      &\times&\times&\times&      &      &      &      &      &      &      &       \\
      &      &      &      &      &      &\times&\times&\times&      &      &      &      &       \\
      &      &      &      &      &      &\times&\times&\times&      &      &      &      &       \\
      &      &      &      &      &      &\times&\times&\times&      &      &      &      &       \\
      &      &      &      &      &      &      &      &      &\times&\times&\times&      &       \\
      &      &      &      &      &      &      &      &      &\times&\times&\times&      &       \\
      &      &      &      &      &      &      &      &      &\times&\times&\times&      &       \\
      &      &      &      &      &      &      &      &      &      &      &      &\times&\times \\
      &      &      &      &      &      &      &      &      &      &      &      &\times&\times 
    \end{array}
  \right) 
\]

\newpage

\subsection{Quantities for the kinematic fit}

The kinematic fit determines the following numbers:

\medskip 
\begin{tabular}[]{lcllcllcll}
$f_{ix-}$         &\h\e&\h $f_1$   &$f_{iy-}$       &\h\e&\h $f_2$   &$f_{iz-}$    &\h\e&\h$ f_3$   &$x$, $y$, and $z$ momentum of incoming $e^-$   \\
$f_{ix+}$         &\h\e&\h $f_4$   &$f_{iy+}$       &\h\e&\h $f_5$   &$f_{iz+}$    &\h\e&\h$ f_6$   &$x$, $y$, and $z$ momentum of incoming $e^+$   \\
$f_{ox-}$         &\h\e&\h $f_7$   &$f_{oy-}$       &\h\e&\h $f_8$   &$f_{oz-}$    &\h\e&\h$ f_9$   &$x$, $y$, and $z$ momentum of outgoing $e^-$   \\
$f_{ox+}$         &\h\e&\h $f_{10}$&$f_{oy+}$       &\h\e&\h $f_{11}$&$f_{oz+}$    &\h\e&\h$ f_{12}$&$x$, $y$, and $z$ momentum of outgoing $e^+$   \\
$\theta_{f\gamma}$&\h\e&\h $f_{13}$&$\phi_{f\gamma}$&\h\e&\h $f_{14}$&$E_{f\gamma}$&\h\e&\h$ h_1   $&$\theta$  and $\phi$, and energy of the photon \\
$\lambda_1$       &    &           &$\lambda_2$     &    &           &$\lambda_3$  &    &           &four Lagrange multipliers for momentum and    \\
$\lambda_4$       &    &           &   	            &    &           &             &    &           &\hskip4mm energy conservation constraints      \\
\end{tabular}
\medskip

\noindent The variables $f_1$ to $f_{14}$ have corresponding measurements.  The variable
$h_1$, the photon energy, is called a ``hidden variable''.  The vector $\bold\alpha$ shall be
defined as a 19-element composite of $\bold f$ (14~elements), $\bold h$ (1~element),
and $\bold\lambda$ (4~elements).

\subsection{Constraints}

We have four constraint equations that have to be satisfied in the kinematic fit:
\[
   \begin{array}{lcl}
      p_{ix-} + p_{ix+} - p_{0x-} - p_{0x+} 
       - E_{f\gamma}\sin{\theta_{f\gamma}}\cos{\phi_{f\gamma}} &=& 0 \qquad\hbox{momentum in $x$}\\
      p_{iy-} + p_{iy+} - p_{0y-} - p_{0y+} 
       - E_{f\gamma}\sin{\theta_{f\gamma}}\sin{\phi_{f\gamma}} &=& 0 \qquad\hbox{momentum in $y$}\\
      p_{iz-} + p_{iz+} - p_{0z-} - p_{0z+} 
       - E_{f\gamma}\cos{\theta_{f\gamma}}                     &=& 0 \qquad\hbox{momentum in $z$}\\
      E_{i-} + E_{i+} - E_{0-} - E_{0+} - E_{f\gamma}          &=& 0 \qquad\hbox{energy}
   \end{array}
\]
Here we use, {\it e.g.},
\[ \begin{array}{rcl}
   E_{i-} &\equiv& \sqrt{ p_{ix-}^2 + p_{iy-}^2 + p_{iz-}^2 + m_e^2 }\\
          &=& \sqrt{ f_1^2     + f_2^2     + f_3^2     + m_e^2 }
   \end{array} \]

\section{The estimated $\chi^2$: $\chi^2_{\rm est}$}

This function is calculated for any given electron-positron-gamma
triplet to determine which triplet should be used for the
kinematic~fit.  At~the end of this subsection, we will have a complete
analytical formula for calculating~$\chi^2_{\rm est}$.

The formula is based on the difference between the initial and final
momentum, $\bold P \equiv \bold P_i - \bold P_o$.  The initial
momentum $\bold P_i$ is the sum of the momenta of the incoming
electron and positron as defined earlier: $\bold P_{i-}$ and $\bold
P_{i+}$.  The measured momenta of the outgoing electron, positron are
given by $\bold P_{0-}$ and $\bold P_{0+}$.

For the outgoing photon, we only have its angles $\theta_{0\gamma}$
and $\phi_{0\gamma}$.  Using the energy constraint \[ E_\gamma =
E_{i-} + E_{i+} - E_{0-} - E_{0+} \] we may substitute the
unknown photon energy $E_\gamma$ with measured values, and we obtain:
\[
  \begin{array}{rcl}
    \bold P_{0\gamma} &\equiv& 
      \left( E_{i-} + E_{i+} - E_{0-} - E_{0+} \right) \left( 
      \begin{array}{c}        
        \sin\theta_{0\gamma}\cos\phi_{0\gamma} \\
        \sin\theta_{0\gamma}\sin\phi_{0\gamma} \\
        \cos\theta_{0\gamma}
      \end{array}
    \right)  \equiv E_\gamma \, \left(
      \begin{array}{c} n_x \\ n_y \\ n_z \end{array} \right) \equiv E_\gamma \,\bold n
  \end{array}
\]
Of course, $\bf n$ is the normal vector, the direction of the photon.

\medskip

Calculating the difference to form vector $\bold P$ is easy:
\[
   \bold P \equiv \left( \begin{array}{c} P_x \\ P_y \\ P_z \end{array} \right) = 
     	 \bold P_{i-} + \bold P_{i+} - \bold
        	   P_{o-} - \bold P_{o+} - \bold P_{o\gamma}
\]
In~the ideal world, this vector would be exactly zero.  
For its error matrix $\bV{p}$, we convert $\bV{\rm all}$, the error matrix of $\bold y$, 
via a transformation matrix $\bold T$ into $\bV{p}$:
\[ \bV p = \bold T^t\,\bV{\rm all}\,\bold T \]
For the transformation matrix $\bold T$ we have to calculate
expressions like $\frac{\partial P_x}{\partial P_{ix-}}$.  We~note
that for $j=x,y,z$:
\[ \frac{\partial (E_\gamma\,n_j)}{\partial P_{ix-}} = \frac{P_{ix-}}{P_{i-}}\,n_j \]
The transformation matrix 
is a $3\times14$ matrix:
\[
   \bold T = \hskip-3pt \left(
       \begin{array}{ccc}
          \dfrac{\partial P_x}{\partial P_{ix-}}& 
          \dfrac{\partial P_y}{\partial P_{ix-}}& 
          \dfrac{\partial P_z}{\partial P_{ix-}}\\*[3mm]
          \dfrac{\partial P_x}{\partial P_{iy-}}& 
          \dfrac{\partial P_y}{\partial P_{iy-}}& 
          \dfrac{\partial P_z}{\partial P_{iy-}}\\*[3mm] 
          \dfrac{\partial P_x}{\partial P_{iz-}}& 
          \dfrac{\partial P_y}{\partial P_{iz-}}& 
          \dfrac{\partial P_z}{\partial P_{iz-}}\\*[3mm] 
          \dfrac{\partial P_x}{\partial P_{ix+}}& 
          \dfrac{\partial P_y}{\partial P_{ix+}}& 
          \dfrac{\partial P_z}{\partial P_{ix+}}\\*[3mm] 
          \dfrac{\partial P_x}{\partial P_{iy+}}& 
          \dfrac{\partial P_y}{\partial P_{iy+}}& 
          \dfrac{\partial P_z}{\partial P_{iy+}}\\*[3mm] 
          \dfrac{\partial P_x}{\partial P_{iz+}}& 
          \dfrac{\partial P_y}{\partial P_{iz+}}& 
          \dfrac{\partial P_z}{\partial P_{iz+}}\\*[3mm] 
          \dfrac{\partial P_x}{\partial P_{0x-}}& 
          \dfrac{\partial P_y}{\partial P_{0x-}}& 
          \dfrac{\partial P_z}{\partial P_{0x-}}\\*[3mm] 
          \dfrac{\partial P_x}{\partial P_{0y-}}& 
          \dfrac{\partial P_y}{\partial P_{0y-}}& 
          \dfrac{\partial P_z}{\partial P_{0y-}}\\*[3mm] 
          \dfrac{\partial P_x}{\partial P_{0z-}}& 
          \dfrac{\partial P_y}{\partial P_{0z-}}& 
          \dfrac{\partial P_z}{\partial P_{0z-}}\\*[3mm] 
          \dfrac{\partial P_x}{\partial P_{0x+}}& 
          \dfrac{\partial P_y}{\partial P_{0x+}}& 
          \dfrac{\partial P_z}{\partial P_{0x+}}\\*[3mm] 
          \dfrac{\partial P_x}{\partial P_{0y+}}& 
          \dfrac{\partial P_y}{\partial P_{0y+}}& 
          \dfrac{\partial P_z}{\partial P_{0y+}}\\*[3mm] 
          \dfrac{\partial P_x}{\partial P_{0z+}}& 
          \dfrac{\partial P_y}{\partial P_{0z+}}& 
          \dfrac{\partial P_z}{\partial P_{0z+}}\\*[3mm] 
          \dfrac{\partial P_x}{\partial \theta_{0\gamma}}& 
          \dfrac{\partial P_y}{\partial \theta_{0\gamma}}& 
          \dfrac{\partial P_z}{\partial \theta_{0\gamma}}\\*[3mm] 
          \dfrac{\partial P_x}{\partial \phi_{0\gamma}}& 
          \dfrac{\partial P_y}{\partial \phi_{0\gamma}}& 
          \dfrac{\partial P_z}{\partial \phi_{0\gamma}} \end{array}
   \right)
   = \left(
      \begin{array}{ccc}
         \p1-\dfrac{P_{ix-}}{E_{i-}}n_x&
        \p\p-\dfrac{P_{ix-}}{E_{i-}}n_y&
        \p\p-\dfrac{P_{ix-}}{E_{i-}}n_z\\*[3mm]
        \p\p-\dfrac{P_{iy-}}{E_{i-}}n_x&
         \p1-\dfrac{P_{iy-}}{E_{i-}}n_y&
        \p\p-\dfrac{P_{iy-}}{E_{i-}}n_z\\*[3mm]
        \p\p-\dfrac{P_{iz-}}{E_{i-}}n_x&
        \p\p-\dfrac{P_{iz-}}{E_{i-}}n_y&
         \p1-\dfrac{P_{iz-}}{E_{i-}}n_z\\*[3mm]
         \p1-\dfrac{P_{ix+}}{E_{i+}}n_x&
        \p\p-\dfrac{P_{ix+}}{E_{i+}}n_y&
        \p\p-\dfrac{P_{ix+}}{E_{i+}}n_z\\*[3mm]
        \p\p-\dfrac{P_{iy+}}{E_{i+}}n_x&
         \p1-\dfrac{P_{iy+}}{E_{i+}}n_y&
        \p\p-\dfrac{P_{iy+}}{E_{i+}}n_z\\*[3mm]
        \p\p-\dfrac{P_{iz+}}{E_{i+}}n_x&
        \p\p-\dfrac{P_{iz+}}{E_{i+}}n_y&
         \p1-\dfrac{P_{iz+}}{E_{i+}}n_z\\*[3mm]
          -1+\dfrac{P_{0x-}}{E_{0-}}n_x&
       \p\p\p\dfrac{P_{0x-}}{E_{0-}}n_y&
       \p\p\p\dfrac{P_{0x-}}{E_{0-}}n_z\\*[3mm]
       \p\p\p\dfrac{P_{0y-}}{E_{0-}}n_x&
          -1+\dfrac{P_{0y-}}{E_{0-}}n_y&
       \p\p\p\dfrac{P_{0y-}}{E_{0-}}n_z\\*[3mm]
       \p\p\p\dfrac{P_{0z-}}{E_{0-}}n_x&
       \p\p\p\dfrac{P_{0z-}}{E_{0-}}n_y&
          -1+\dfrac{P_{0z-}}{E_{0-}}n_z\\*[3mm]
          -1+\dfrac{P_{0x+}}{E_{0+}}n_x&
       \p\p\p\dfrac{P_{0x+}}{E_{0+}}n_y&
       \p\p\p\dfrac{P_{0x+}}{E_{0+}}n_z\\*[3mm]
       \p\p\p\dfrac{P_{0y+}}{E_{0+}}n_x&
          -1+\dfrac{P_{0y+}}{E_{0+}}n_y&
       \p\p\p\dfrac{P_{0y+}}{E_{0+}}n_z\\*[3mm]
       \p\p\p\dfrac{P_{0z+}}{E_{0+}}n_x&
       \p\p\p\dfrac{P_{0z+}}{E_{0+}}n_y&
          -1+\dfrac{P_{0z+}}{E_{0+}}n_z\\*[3mm]
        - E_\gamma\cos\theta_{0\gamma}\cos\phi_{0\gamma}& 
        - E_\gamma\cos\theta_{0\gamma}\sin\phi_{0\gamma}& 
       \p E_\gamma\sin\theta_{0\gamma}\\*[3mm] 
       \p E_\gamma\sin\theta_{0\gamma}\sin\phi_{0\gamma}& 
        - E_\gamma\sin\theta_{0\gamma}\cos\phi_{0\gamma}& 
          0  
       \end{array}
    \right)
\]

Now we have $\BV p = \bold T^t\,\BV{\rm all}\,\bold T$, and hence we
may calculate~$\chi^2_{\rm est}$:
 \[ \chi^2_{\rm est} = \bold P^t\,\BV p\,\bold P \]

What is the meaning of this $\chi^2$?  We can say that the 14 input
variables are used to measure $\bold P$, and $\chi^2_{\rm est}$ tells us the
deviation of the measured $\bold P$ from the expected $\bold P$, which
is zero.

\section{The kinematic fit}

For the derivation of the kinematic fit algorithm, we follow the description of
Louis Lyons, page 151, 152~\cite{bib:lyons}.

\subsection{The $\chi^2$-Function}

The real $\chi^2$-function can be written down in the following way:

\[
 \begin{array}{rcl}
 \chi^2 &=& (\bold f-\bold m)^t \BV{\rm all} (\bold f-\bold m) \\
        &+& \lambda_1 \left[p_{xi-}+p_{xi+}-p_{xo-}-p_{xo+}-E_\gamma\sin\theta_\gamma\cos\phi_\gamma\right] \\
        &+& \lambda_2 \left[p_{yi-}+p_{yi+}-p_{yo-}-p_{yo+}-E_\gamma\sin\theta_\gamma\sin\phi_\gamma\right] \\
        &+& \lambda_3 \left[p_{zi-}+p_{zi+}-p_{zo-}-p_{zo+}-E_\gamma\cos\theta_\gamma\right] \\
    	&+& \lambda_4 \left[E_{i-} +E_{i+} -E_{o-} -E_{o+} -E_\gamma\right] 
 \end{array}
\]

The constraint equations are here included via Lagrange multipliers.
To~minimize this $\chi^2$, we~could use a standard package like {\tt
MINUIT}, but standard packages are always slower than specially
adapted code.  Since the $\chi^2$-minimization is being done millions
of times, it pays off to write special code for the minimization.  In
addition, {\tt MINUIT} is not supported in \BaBar's Online Prompt
Reconstruction.

\subsection{Derivation of kinematic fit algorithm}

At the minimum of $\chi^2$, its first derivatives are to be zero.
Lyons uses for this the following equations:
\begin{eqnarray*} 
  \frac{\partial\chi^2}{\partial\al_i}     &=& 0 \qquad\hbox{for $i=1$ to 14} \\
  \frac{\partial\chi^2}{\partial h}        &=& 0 \qquad\hbox{here $h=E_\gamma = \al_{15}$} \\
  \frac{\partial\chi^2}{\partial\lambda_k} &=& 0 \qquad\hbox{here $\lambda_1 = \al_{16}$ etc.} 
\end{eqnarray*}

\noindent The three equation sets can be written as:\footnote{The
factor~2 in front of $\BV{\rm all}$ is missing in Lyons'
book~\cite{bib:lyons}.  We~could easily remove this factor from our
formulas by re-defining the Lagrange multipliers in the
$\chi^2$-function with a factor~2.  This would not change the fit
result or errors, as long as the subsequent calculations were carried
out consistently.}
\begin{eqnarray*} 
  2\,\bold G\,(\bold f-\bold m) + \bold D^t\,\bold\lambda &=& \bold 0 \\
  \bold E^t \, \bold\lambda &=& \bold 0 \\
  \bold C &=& \bold 0 
\end{eqnarray*}
where $\bold G$ is the $14\times14$ inverse error matrix of the
measurements which we also call $\BV{\rm all}$.
\begin{eqnarray*} \bold D = \pmatrix{
    \sfrac{\partial C_1}{\partial\al_1}&\ldots&\sfrac{\partial C_1}{\partial\al_{14}} \cr
    \sfrac{\partial C_2}{\partial\al_1}&\ldots&\sfrac{\partial C_2}{\partial\al_{14}} \cr
    \sfrac{\partial C_3}{\partial\al_1}&\ldots&\sfrac{\partial C_3}{\partial\al_{14}} \cr
    \sfrac{\partial C_4}{\partial\al_1}&\ldots&\sfrac{\partial C_4}{\partial\al_{14}} } 
\qquad\hbox{and}\qquad
  \bold E = \pmatrix{
    \sfrac{\partial C_1}{\partial\al_{15}}\cr
    \sfrac{\partial C_2}{\partial\al_{15}}\cr
    \sfrac{\partial C_3}{\partial\al_{15}}\cr
    \sfrac{\partial C_4}{\partial\al_{15}}}
\end{eqnarray*}

\noindent We now expand the constraint equations $\bold C$ around $f_0$ and $h_0$, and 
we obtain for the four equations $C_k$ with $k=1$ to~4:
\begin{eqnarray} C_k \approx C_k^{(0)} 
   + \sum_{i=1}^{14}\,\frac{\partial C^{(0)}}{\partial f_i}\,(f_i-f_i^{(0)}) 
   + \frac{\partial C^{(0)}}{\partial h}\,(h-h^{(0)}) = 0 
\label{eq:constraintExpansion}\end{eqnarray}

\noindent We may rewrite this into:
\[ \sum_{i=1}^{14}\,\frac{\partial C^{(0)}}{\partial f_i}\,(f_i-m_i) 
   + \frac{\partial C^{(0)}}{\partial h}\,(h-h^{(0)}) = -C_k^{(0)} +
  \sum_{i=1}^{14}\,\frac{\partial C^{(0)}}{\partial f_i}\,(f_i^{(0)}-m_i^{(0)})  \]

\noindent Now we collect everything, use the definitions for $\bold M$, $\bold Y$, and $\bold Z$,
\[   \bold M = \left( \begin{array}{ccc} 2\bold G & \bold 0 & \bold D^t \\
                                          \bold 0 & \bold 0 & \bold E^t \\
                                          \bold D & \bold E & \bold 0   \end{array} \right) \qquad
     \bold Y = \left( \begin{array}{c} \bold f-\bold m \\ \bold h - \bold h_0 \\ \bold \lambda  \end{array} \right) \qquad
     \bold Z = \left( \begin{array}{c} \bold 0 \\ \bold 0 \\ -\bold R  \end{array} \right) \]
with 
\begin{eqnarray*} \bold R = \bold C^{(0)} - \bold D(\bold f_0-\bold m) 
 = \pmatrix{ C_1(\bold f_0,\bold h_0) \cr 
             C_2(\bold f_0,\bold h_0) \cr 
             C_3(\bold f_0,\bold h_0) \cr 
             C_4(\bold f_0,\bold h_0) } - \bold D (\bold f_0 - \bold m),    \end{eqnarray*}
and we see:
\[ \bold M\,\bold Y = \bold Z \]
This is the equation we have to solve.  Since the constraint equations
${\bold C}={\bold 0}$ contain non-linear functions like $\sin\theta_\gamma$,
Eq.~(\ref{eq:constraintExpansion}) is only an approximation, and we
have to iterate as described in the next section.

\subsection{Recipe for the kinematic fit algorithm}

The matrix $\bold M$ and the vector $\bold Z$ are functions of the
measurements and their error matrices as well as of the parameters
$\bold\alpha$.  The vector $\bold Y$ is, as mentioned above,
\[ \bold Y = \pmatrix{ \bold f - \bold m \cr \bold h - \bold h_0 \cr \bold\lambda } \] 
and can be calculated with:
\[  \bold Y=\bold M^{-1}\,\bold Z \]

Here is the iteration: Initially, we will use for the fit quantities
$\bold f_0 = \bold m$, {\it i.e.\null} the measured quantities.  For
$\bold h = h_0$, we calculate the photon energy via simple energy
conservation.  These together with the measured quantities allow us to
calculate $\bold M$ and $\bold Z$.  We multiply the inverse of $\bold
M$ with $\bold Z$ and obtain $\bold Y$.  This result will then give us
a better set of $\bold f$ and $\bold h$, which we again use to
calculate $\bold M$ and $\bold Z$, and then a better $\bold Y$.  And
we continue until our constraint equations are sufficiently fulfilled
and the quantities $\bold f$ and $\bold h$ are stable.

It might happen that the iteration does not converge at the minimum,
but wanders off into unphysical numbers.  In~that case, it would be
good to have a certain boundary box around the point.  If~the step
would make the point lie outside the box, then the program would
change the step so that the point would be back inside.  It might be
good to implement this, although the radiative Bhabha fitting does not
seem to need this part of the algorithm.

\subsection{Details of matrices and vectors used in the kinematic fit}

We~define the following variables:
\[
   E_i = \left\{ \begin{array}{ll} 
    E_{i-} &\qquad\hbox{for $i=1$,   2,  3 \ ($p_{ix-}$, $p_{iy-}$, $p_{iz-}$)}\\
    E_{i+} &\qquad\hbox{for $i=4$,   5,  6 \ ($p_{ix+}$, $p_{iy+}$, $p_{iz+}$)}\\
    E_{0-} &\qquad\hbox{for $i=7$,   8,  9 \ ($p_{0x-}$, $p_{0y-}$, $p_{0z-}$)}\\
    E_{0+} &\qquad\hbox{for $i=10$, 11, 12 \ ($p_{0x+}$, $p_{0y+}$, $p_{0z+}$)}
   \end{array}\right.   
\]

\[
   s_i = \left\{ \begin{array}{rl}
     1 & \qquad\hbox{for $i\le6$ \ ($p_{ix-}$, $p_{iy-}$, $p_{iz-}$, $p_{ix+}$, $p_{iy+}$, $p_{iz+}$)}\\             
    -1 & \qquad\hbox{for $i>6$   \ ($p_{0x-}$, $p_{0y-}$, $p_{0z-}$, $p_{0x+}$, $p_{0y+}$, $p_{0z+}$)}           
   \end{array} \right.
\]

For $4\times14$ matrix $\bold D$ we need the following expressions:

Row $j=1$ to 3, columns $i=1$ to 12:
\[ \frac{\partial C_j}{\partial\al_i} = \left\{ 
    \begin{array}{cl} 
       s_i &\hbox{if $i=j$ or $i=j+3$ or $i=j+6$ or $i=j+9$} \\
        0  &\hbox{else} 
    \end{array} \right.\]

Row $j=1$, column $i=13$:
\[ \frac{\partial C_1}{\partial\al_{13}} = - E_{f\gamma}\,\cos{\theta_{f\gamma}}\,\cos{\phi_{f\gamma}} 
                                          = - \al_{15}   \,\cos{\al_{13}}        \,\cos{\al_{14}}       \]

Row $j=1$, column $i=14$:
\[ \frac{\partial C_1}{\partial\al_{14}} =   E_{f\gamma}\,\sin{\theta_{f\gamma}}\,\sin{\phi_{f\gamma}} 
                                          =   \al_{15}   \,\sin{\al_{13}}        \,\sin{\al_{14}}       \]

Row $j=2$, column $i=13$:
\[ \frac{\partial C_2}{\partial\al_{13}} = - E_{f\gamma}\,\cos{\theta_{f\gamma}}\,\sin{\phi_{f\gamma}} 
                                          = - \al_{15}   \,\cos{\al_{13}}        \,\sin{\al_{14}}       \]

Row $j=2$, column $i=14$:
\[ \frac{\partial C_2}{\partial\al_{14}} = - E_{f\gamma}\,\sin{\theta_{f\gamma}}\,\cos{\phi_{f\gamma}} 
                                          = - \al_{15}   \,\sin{\al_{13}}        \,\cos{\al_{14}}       \]

Row $j=3$, column $i=13$:
\[ \frac{\partial C_3}{\partial\al_{13}} =   E_{f\gamma}\,\sin{\theta_{f\gamma}} = \al_{15}\,\sin{\al_{13}} \]
										
Row $j=3$, column $i=14$:							
\[ \frac{\partial C_3}{\partial\al_{14}} = 0 \]

Row $j=4$, columns $i=1$ to 12:
\[ \frac{\partial C_4}{\partial\al_i} = s_i\frac{\al_i}{E_i} \]

Row $j=4$, column $i=13$ and 14:
\[ \frac{\partial C_4}{\partial\al_{13}} = 
   \frac{\partial C_4}{\partial\al_{14}} = 0 \]

The $4\times1$ matrix $\bold E$ is:
\[ \bold E = \left(\begin{array}{c} 
   -\sin{\theta_{f\gamma}}\,\cos{\phi_{f\gamma}} \\
   -\sin{\theta_{f\gamma}}\,\sin{\phi_{f\gamma}} \\
   -\cos{\theta_{f\gamma}} \\
   -1 
   \end{array}\right) = 
   \left(\begin{array}{c} 
   -\sin{\al_{13}}\,\cos{\al_{14}} \\
   -\sin{\al_{13}}\,\sin{\al_{14}} \\
   -\cos{\al_{13}} \\
   -1 
   \end{array}\right) \]

\subsection{The error matrix of the fit}

The second partial derivatives of $\chi^2$ appear in the error matrix 
of the fit parameters:
\begin{eqnarray*} \bold H = \left( \frac{1}{2} 
  \frac{\partial^2\chi^2}{\partial\alpha_i\partial\alpha_j}\right)^{-1} 
\end{eqnarray*}
So~in our case, $\bold H$ is a $19\times19$ matrix.  The detailed expressions for 
the second derivatives of $\chi^2$ will be given in the following section.

\subsection{Tests for goodness of fit}

After completing the iteration on the kinematic fit, one wants to make
sure that all quantities are indeed correct.

Besides the obvious tests that the constraint equations are satisfied,
one can check that indeed a minimum was reached.  For this, one may
wiggle each final value $\alpha_1$ to $\alpha_{14}$ and
recalculate~$\chi^2$.  In~our case we have in the $\chi^2$-function
the terms with the Lagrange multipliers.  Just recalculating the
$\chi^2$ function will not lead to correct results, since the found
vector $\bold\alpha$ is a minimum only when also requiring the
constraints.  So one has to redo the fit while forcing the selected
element of $\bold\alpha$ to the off-minimum~value.

This wiggling allows us to map out the minimum, and it also tells us
whether the fit error returned for that parameter is reasonable.  
If we fix $E_{f\gamma}$ to be $\pm1\sigma_{\gamma\,{\rm fit}}$ way
from the real fit result, then the $\chi^2$ should rise by~1 in either
direction.  When mapping out this rise, one will see the shape of a
parabola.  When the formulas are complicated and/or one is far away from the 
minimum, the parabola will be distorted.  

In our case, we can indeed calculate the fit error for $E_{f\gamma}$,
but if this would be impossible, one can find the fit error by mapping
out the minimum with the above described re-fitting with fixed values.
The~$\pm1\sigma$-error is then defined to be where $\chi^2$ is 1~unit
above the minimum.  As mentioned, this function may be distorted when
far away from the minimum.  A~complicated $\chi^2$-function might even
distort the $\pm1\sigma$-area.  In this case, one can take the minimum
and two points very close to it, fit a parabola through these three
points, and take the sigma from that parabola as the error.

The same process also works for the hidden parameter (fitted photon
energy), and we definitely have to re-fit since the fitted photon
energy only appears in the constraints, where the Lagrange multipliers
would influence the outcome.

Here is how we have to modify the formulas for re-fitting:

\subsubsection{Re-fitting with fixed $E_{f\gamma}$}

We~want to redo the fit with the photon energy fixed to $E_{\rm fix} =
E_{f\gamma}+\epsilon$.  To the $\chi^2$-function, we add the term
\begin{eqnarray*} + \quad X \left(E_\gamma-E_{\rm fix}\right)^2 
 \end{eqnarray*} where $X$ is a large number compared to the original
$\chi^2$.  If we now minimize this new $\chi^2$-function, the
additional term adds a large penalty to any deviation of $E_\gamma$
from $E_{\rm fix}$.

Going through the derivation again, we find the following places
that have to be changed in the code:

$\bullet$ First partial derivative $\frac{\partial\chi^2}{\partial
\alpha_i}$ for $i=15$ [for $i=k$] has the additional term
``$+2X(E_\gamma-E_{\rm fix})$''.

$\bullet$ No change to second partial derivatives.

$\bullet$ Matrix $\bf M$ has the additional term ``$+2X$'' at (15,15).
This means that the (15,15)-element of $\bold M$ is no longer zero.

$\bullet$ Vector $\bold Z$ has an additional term at position 15:
\[ \bold Z = \pmatrix{ \bold 0 \cr -2X(h_0-E_{\rm fix}) \cr -\bold R  } \].

These are all necessary changes.  The iteration should converge again,
but this time always result in $E_\gamma=E_{\rm fix}$ for sufficiently
large~$X$.

\subsubsection{Re-fitting with fixed $f_{k}$}

Let us now wiggle one of the measurement variables $\alpha_1$ to
$\alpha_{14}$.  When fixing $f_k$ to $f_k=f_{k\,{\rm fix}}$, we add
the term
\begin{eqnarray*} + \quad X \left(f_k-f_{k\,{\rm fix}}\right)^2  \end{eqnarray*} 
to the $\chi^2$-function.  Again, $X$ is a large number compared to
the original $\chi^2$.  The following changes have to be made in the
formulas of the algorithm:

$\bullet$ The first partial derivative
$\partial\chi^2/\partial\alpha_i$ gets for $i=k$ the additional term
``$+2X(f_k-f_{k\,{\rm fix}})$''.

$\bullet$ Again no change to second partial derivatives.

$\bullet$ Matrix $\bf M$ gets at position $(k,k)$ the additional
term~``$+2X$''.

$\bullet$ Vector $\bf Z$ has at position $k$ the entry
``$-2X(m_k-f_{k\,{\rm fix}})$''.

\subsubsection{Confidence Level}

If~all errors of the measurements are nicely described by Gaussian
distributions, and if all events are what we think they are, {\it
i.e.}, (in our case) radiative Bhabhas, then the $\chi^2$ values of
the fits should be distributed like the $\chi^2$-distribution for
$n=3$ (3 because out fit is a 3-constraint fit).  Instead of looking
at the $\chi^2$ distributions directly, it is easier to map the
$\chi^2$ to a flat distribution with values between 0 and 1.  This
value is then called the confidence level~(C.L.) of the event.  If the
$\chi^2$ is really distributed as it should be, the confidence level
will have a flat distribution.

So we are looking for two things in the C.L. distribution: 

(1) Most of the region should have a flat distribution.  If not, the
errors used in the fit might be too large or too small.  If the errors
are underestimated, the $\chi^2$ will be larger than expected, and the
confidence level distribution will be tilted downward (when going from~0 to~1).
Vice-versa, if the errors are overestimated, the C.L. distribution
will be tilted upward.  More information on the validity of errors
might be obtained from the ``pull'' distributions described later.

(2) A peak at zero indicates events that do not fulfill the kinematics
of radiative Bhabhas at~all.  They will result in very large $\chi^2$
(=very small C.L., close to zero).  These events can come from
backgrounds or misidentified tracks.  What can we do?  We can improve
our selection criteria.  Or we can cut out all events belonging to
that peak, taking only those events that are part of the flat
distribution.  A~cut on the confidence level is, of course, equivalent
to a cut on~$\chi^2$.

\subsubsection{The ``Pull''}

For each measured variable, one can plot the so-called
``pull''~\cite{bib:eadie} or ``normalized stretch
values''~\cite{bib:blobel}~\cite{bib:roe}:
\begin{eqnarray*} 
  \hbox{pull \null}p = \frac{\hbox{meas} - \hbox{fit}}{\sqrt{\sigma_{\rm meas}-\sigma_{\rm fit}}}    
\end{eqnarray*}
The minus sign in the square root comes from the strong correlation
between the measured and the fitted quantity, and ``{\it still puzzles
many users\/}''~\cite{bib:eadie}. If~all measured errors were estimated
correctly and the conditions for the fit were satisfied ({\it
e.g.}\null, the event was really a radiative Bhabha event), then the
pull quantity will be distributed like a Gaussian centered at~0 with
$\sigma=1$.  If an error is for example overestimated, the pull
quantity will have a more narrow distribution.  In this case, the confidence
level should also be affected, displaying a tilt in its distribution.  

To check whether a systematic increase or decrease of one or more
errors would improve the pull and/or the confidence level
distributions, one can redo the whole analysis with increased or
decreased errors.  Perhaps one can find a set of corrections that
create nice pull distributions and a nice confidence level
distribution.  If the errors are really not correct, one should talk
with the colleagues who are responsible for the errors.  However,
abnormal pull quantities might not be always created by incorrect
errors.  Systematically shifted measurements could also cause such
symptoms.

\section{$\chi^2$-Function --- First Derivatives}

For this set of equations, we will use the following notation:
\[
   L_i = \left\{ \begin{array}{rl}
    \lambda_1   & \qquad\hbox{for $i=1$, 4, 7, 10 \ ($p_{ix-}$, $p_{ix+}$, $p_{0x-}$, $p_{0x+}$)}\\
    \lambda_2   & \qquad\hbox{for $i=2$, 5, 8, 11 \ ($p_{iy-}$, $p_{iy+}$, $p_{0y-}$, $p_{0y+}$)}\\
    \lambda_3   & \qquad\hbox{for $i=3$, 6, 9, 12 \ ($p_{iz-}$, $p_{iz+}$, $p_{0z-}$, $p_{0z+}$)}
   \end{array} \right.
\]

Now we calculate the first partial derivatives of the $\chi^2$-function, i.e., the
19 equations $\partial \chi^2/\partial \alpha_i$.

\noindent For $i=1$ to 12:
\[ \frac{\partial\chi^2}{\partial\alpha_i} = 2\,\sum_{j=1}^{14} \BV{{\rm all}\,ij} (f_j-m_j) 
 	+ s_iL_i + s_i\lambda_4\frac{f_i}{E_i}
\]

\noindent For $i=13$:
\begin{eqnarray*}
    \frac{\partial\chi^2}{\partial\alpha_i} 
      &=& 2\,\sum_{j=1}^{14} \BV{{\rm all}\,ij} (f_j-m_j) 
          - \lambda_1 E_\gamma\cos{\theta_\gamma}\cos{\phi_\gamma}
          - \lambda_2 E_\gamma\cos{\theta_\gamma}\sin{\phi_\gamma}
          + \lambda_3 E_\gamma\sin{\theta_\gamma}\\
      &=& 2\,\sum_{j=1}^{14} \BV{{\rm all}\,ij} (\al_j-m_j)\\
      &&  - \al_{16}\al_{15}\cos{\al_{13}}\cos{\al_{14}}
          - \al_{17}\al_{15}\cos{\al_{13}}\sin{\al_{14}}
          + \al_{18}\al_{15}\sin{\al_{13}} \end{eqnarray*}

\noindent For $i=14$:
\begin{eqnarray*}
   \frac{\partial\chi^2}{\partial\alpha_i} &=&
 	 2\,\sum_{j=1}^{14} \BV{{\rm all}\,ij} (f_j-m_j) 
              + \lambda_1 E_\gamma\sin{\theta_\gamma}\sin{\phi_\gamma}
              - \lambda_2 E_\gamma\sin{\theta_\gamma}\cos{\phi_\gamma}\\
     &=& 2\,\sum_{j=1}^{14} \BV{{\rm all}\,ij} (\al_j-m_j) 
              + \al_{16}\al_{15}\sin{\al_{13}}\sin{\al_{14}}
              - \al_{17}\al_{15}\sin{\al_{13}}\cos{\al_{14}} \end{eqnarray*}

\noindent For $i=15$:
\begin{eqnarray*}
   \frac{\partial\chi^2}{\partial\alpha_i} &=& 
              - \lambda_1 \sin{\theta_\gamma}\cos{\phi_\gamma}
              - \lambda_2 \sin{\theta_\gamma}\sin{\phi_\gamma}
              - \lambda_3 \cos{\theta_\gamma}
              - \lambda_4\\
     &=&      - \al_{16} \sin{\al_{13}}\cos{\al_{14}}
              - \al_{17} \sin{\al_{13}}\sin{\al_{14}}
              - \al_{18} \cos{\al_{13}}
              - \al_{19} \end{eqnarray*}

\noindent For $i=16$:
\begin{eqnarray*}
   \frac{\partial\chi^2}{\partial\alpha_i} 
   &=& p_{xi-}+p_{xi+}-p_{x0-}-p_{x0+}-E_\gamma\sin\theta_\gamma\cos\phi_\gamma \\
   &=& \al_{1}+\al_{4}-\al_{7}-\al_{10}-\al_{15}\sin \al_{13}\cos \al_{14} \end{eqnarray*}

\noindent For $i=17$:
\begin{eqnarray*}
   \frac{\partial\chi^2}{\partial\alpha_i} 
   &=& p_{yi-}+p_{yi+}-p_{y0-}-p_{y0+}-E_\gamma\sin\theta_\gamma\sin\phi_\gamma \\ 
   &=& \al_{2}+\al_{5}-\al_{8}-\al_{11}-\al_{15}\sin \al_{13}\sin \al_{14}      \end{eqnarray*}

\noindent For $i=18$:
\begin{eqnarray*}
   \frac{\partial\chi^2}{\partial\alpha_i}
   &=& p_{zi-}+p_{zi+}-p_{z0-}-p_{z0+}-E_\gamma\cos\theta_\gamma \\
   &=& \al_{3}+\al_{6}-\al_{9}-\al_{12}-\al_{15}\cos{\al_{13}}   \end{eqnarray*}

\noindent For $i=19$:
\begin{eqnarray*}
   \frac{\partial\chi^2}{\partial\alpha_i}
   &=& E_{i-} +E_{i+} -E_{0-} -E_{0+} -E_\gamma	\\
   &=& E_{i-} +E_{i+} -E_{0-} -E_{0+} -\al_{15} \end{eqnarray*}

\section{$\chi^2$-Function --- Second Derivatives}

\noindent For $i=1$ to 12 and $j=1$ to 12:
\[ \begin{array}{rclcl}
  	\dfrac{\partial^2\chi^2}{\partial\alpha_j\partial\alpha_i} 
	&=& 2\,\BV{{\rm all}\,ij} + s_i\lambda_4\,\frac{E_i-f_i^2/E_i}{E_i^2}
  	&=& 2\,\BV{{\rm all}\,ij} - s_i\al_{19} \,\frac{\al_i^2-E_i^2}{E_i^3}
        \qquad\hbox{if $i=j$} \\*[3mm]
        &=& 2\,\BV{{\rm all}\,ij} - s_i\lambda_4\,\frac{f_i\,f_j}    {E_i^3}
        &=& 2\,\BV{{\rm all}\,ij} - s_i\al_{19} \,\frac{\al_i\,\al_j}{E_i^3}
        \qquad\hbox{if $E_i=E_j$ by definition} \\*[3mm]
        &=& 2\,\BV{{\rm all}\,ij} & & 
        \qquad\hbox{else} \\*[3mm]
   \end{array} \]

\noindent For $i=1$ to 12 and $j=13$ to 14:
\[ \frac{\partial^2\chi^2}{\partial\alpha_j\partial\alpha_i} = 2\,\BV{{\rm all}\,ij} \]

\noindent For $i=1$ to 12 and $j=15$:
\[ \frac{\partial^2\chi^2}{\partial\alpha_j\partial\alpha_i} = 0 \]

\noindent For $i=1$ to 12 and $j=16$ to 18:
 \begin{eqnarray*}
        \frac{\partial^2\chi^2}{\partial\alpha_j\partial\alpha_i} 
        &=& \hbox{\rlap{$s_i$}} \qquad \hbox{if $L_i=L_j$ by definition}\\
        &=& \hbox{\rlap{0}}     \qquad \hbox{else}              \end{eqnarray*} 

\noindent For $i=1$ to 12 and $j=19$:
\[  \frac{\partial^2\chi^2}{\partial\alpha_j\partial\alpha_i} = 
	s_i\frac{f_i}{E_i}= s_i\frac{\al_i}{E_i} \]

\noindent For $i=13$ and $j=13$:
\begin{eqnarray*} 
	\frac{\partial^2\chi^2}{\partial\alpha_j\partial\alpha_i} 
	&=& 2\,\BV{{\rm all}\,ij} 
          + \lambda_1 E_\gamma\sin{\theta_\gamma}\cos{\phi_\gamma}
          + \lambda_2 E_\gamma\sin{\theta_\gamma}\sin{\phi_\gamma}
          + \lambda_3 E_\gamma\cos{\theta_\gamma} \\
	&=& 2\,\BV{{\rm all}\,ij} 
          + \al_{16} \al_{15}\sin{\al_{13}}\cos{\al_{14}}
          + \al_{17} \al_{15}\sin{\al_{13}}\sin{\al_{14}}
          + \al_{18} \al_{15}\cos{\al_{13}} \end{eqnarray*}

\noindent For $i=13$ and $j=14$:
\begin{eqnarray*} 
	\frac{\partial^2\chi^2}{\partial\alpha_j\partial\alpha_i} 
	&=& 2\,\BV{{\rm all}\,ij} 
          + \lambda_1 E_\gamma\cos{\theta_\gamma}\sin{\phi_\gamma}
          - \lambda_2 E_\gamma\cos{\theta_\gamma}\cos{\phi_\gamma} \\
	&=& 2\,\BV{{\rm all}\,ij} 
          + \al_{16}\al_{15}\cos{\al_{13}}\sin{\al_{14}}
          - \al_{17}\al_{15}\cos{\al_{13}}\cos{\al_{14}} \end{eqnarray*}

\noindent For $i=13$ and $j=15$:
\begin{eqnarray*}
	\frac{\partial^2\chi^2}{\partial\alpha_j\partial\alpha_i} 
	&=& - \lambda_1 \cos{\theta_\gamma}\cos{\phi_\gamma}
            - \lambda_2 \cos{\theta_\gamma}\sin{\phi_\gamma}
            + \lambda_3 \sin{\theta_\gamma} \\
	&=& - \al_{16}\cos{\al_{13}}\cos{\al_{14}}
            - \al_{17}\cos{\al_{13}}\sin{\al_{14}}
            + \al_{18}\sin{\al_{13}} \end{eqnarray*}

\noindent For $i=13$ and $j=16$:
\[    \frac{\partial^2\chi^2}{\partial\alpha_j\partial\alpha_i} 
	 = - E_\gamma\cos{\theta_\gamma}\cos{\phi_\gamma}     	   
	 = - \al_{15}\cos{\al_{13}}\cos{\al_{14}} \]

\noindent For $i=13$ and $j=17$:
\[      \frac{\partial^2\chi^2}{\partial\alpha_j\partial\alpha_i} 
	= - E_\gamma\cos{\theta_\gamma}\sin{\phi_\gamma} 
	= - \al_{15}\cos{\al_{13}}\sin{\al_{14}} \]

\noindent For $i=13$ and $j=18$:
\[  \frac{\partial^2\chi^2}{\partial\alpha_j\partial\alpha_i} 
	= E_\gamma\sin{\theta_\gamma} 
	= \al_{15}\sin{\al_{13}} \]

\noindent For $i=13$ and $j=19$:
\[  \frac{\partial^2\chi^2}{\partial\alpha_j\partial\alpha_i} = 0 \]

\noindent For $i=14$ and $j=14$:
\begin{eqnarray*}
	\dfrac{\partial^2\chi^2}{\partial\alpha_j\partial\alpha_i} 
	&=& 2\,\BV{{\rm all}\,ij}  
              + \lambda_1 E_\gamma\sin{\theta_\gamma}\cos{\phi_\gamma}
              + \lambda_2 E_\gamma\sin{\theta_\gamma}\sin{\phi_\gamma} \\
	&=& 2\,\BV{{\rm all}\,ij}  
              + \al_{16} \al_{15}\sin{\al_{13}}\cos{\al_{14}}
              + \al_{17} \al_{15}\sin{\al_{13}}\sin{\al_{14}} \end{eqnarray*}

\noindent For $i=14$ and $j=15$:
\begin{eqnarray*}
	\frac{\partial^2\chi^2}{\partial\alpha_j\partial\alpha_i} 
	&=& \lambda_1 \sin{\theta_\gamma}\sin{\phi_\gamma}
              - \lambda_2 \sin{\theta_\gamma}\cos{\phi_\gamma} \\
	&=& \al_{16} \sin{\al_{13}}\sin{\al_{14}}
          - \al_{17} \sin{\al_{13}}\cos{\al_{14}} \end{eqnarray*}

\noindent For $i=14$ and $j=16$:
\begin{eqnarray*}
	\frac{\partial^2\chi^2}{\partial\alpha_j\partial\alpha_i} 
	&=& E_\gamma\sin{\theta_\gamma}\sin{\phi_\gamma} \\
	&=& \al_{15}\sin{\al_{13}}\sin{\al_{14}}  \end{eqnarray*}

\noindent For $i=14$ and $j=17$:
\[  \begin{array}{rcl}
	\frac{\partial^2\chi^2}{\partial\alpha_j\partial\alpha_i} 
	&=& - E_\gamma\sin{\theta_\gamma}\cos{\phi_\gamma} \\*[3mm]
	&=& - \al_{15}\sin{\al_{13}}\cos{\al_{14}} \end{array} \]

\noindent For $i=14$ and $j=18$ and 19:
\[  \frac{\partial^2\chi^2}{\partial\alpha_j\partial\alpha_i} = 0 \]

\noindent For $i=15$ and $j=15$:
\[  \frac{\partial^2\chi^2}{\partial\alpha_j\partial\alpha_i} = 0 \]

\noindent For $i=15$ and $j=16$:
\[  \frac{\partial^2\chi^2}{\partial\alpha_j\partial\alpha_i} 
	= - \sin{\theta_\gamma}\cos{\phi_\gamma} 
	= - \sin{\al_{13}}\cos{\al_{14}} \]

\noindent For $i=15$ and $j=17$:
\[  \frac{\partial^2\chi^2}{\partial\alpha_j\partial\alpha_i} 
	= - \sin{\theta_\gamma}\sin{\phi_\gamma} 
	= - \sin{\al_{13}}\sin{\al_{14}} \]

\noindent For $i=15$ and $j=18$:
\[  \frac{\partial^2\chi^2}{\partial\alpha_j\partial\alpha_i} 
	= - \cos{\theta_\gamma} = - \cos{\al_{13}} \]

\noindent For $i=15$ and $j=19$:
\[  \frac{\partial^2\chi^2}{\partial\alpha_j\partial\alpha_i} = - 1 \]

\noindent For $i=16$ to 19 and $j=16$ to 19:
$$  \frac{\partial^2\chi^2}{\partial\alpha_j\partial\alpha_i} = 0 $$

\section*{Acknowledgments}

I~thank Bill Dunwoodie (SLAC) for his patience with my numerous
questions.  His help and expertise were invaluable in understanding
all issues on the estimated~$\chi^2$ and the kinematic fit.

\end{document}